\documentclass[fleqn,usenatbib]{mnras}

\usepackage{array}

\usepackage[T1]{fontenc}
\DeclareRobustCommand{\VAN}[3]{#2}
\let\VANthebibliography\thebibliography
\def\thebibliography{\DeclareRobustCommand{\VAN}[3]{##3}\VANthebibliography}

\usepackage{graphicx}	
\usepackage{amsmath}	
\usepackage{amssymb}	

\usepackage{newtxtext,newtxmath}
\usepackage[table]{xcolor}

\title[The Kormendy relation of cluster galaxies]{The Kormendy relation of cluster galaxies in  PPS regions}

\author[Ribeiro et al.]{
A.L.B. Ribeiro$^{1}$, \thanks{E-mail: albr@uesc.br}
P.A.A. Lopes$^{2}$,
D.F. Morell$^{3}$,
C.C. Dantas$^{4}$,
M.H.S. Fonseca$^{2}$,\and
B.G. Amarante$^{1}$,
F.R. Morais-Neto$^{1}$
\\
\\
$^{1}$Laborat\'orio de Astrof\'isica Te\'orica e Observacional, Universidade Estadual de Santa Cruz, Ilh\'eus, BA 454650-000, Brazil\\
$^{2}$Observat\'orio do Valongo, Universidade Federal do Rio de Janeiro, RJ 20080-090, Brazil\\
$^{3}$Instituto de Astronomia, Universidad Nacional Autónoma de México, Apartado Postal 70-264, Ciudad de México, D.F., México\\
$^{4}$Instituto Nacional de Pesquisas Espaciais/MCTI, SP 12227-900, Brazil\\
}

\date{Accepted XXX. Received YYY; in original form ZZZ}

\pubyear{2015}

\begin{document}
\label{firstpage}
\pagerange{\pageref{firstpage}--\pageref{lastpage}}
\maketitle

\begin{abstract}
We study a sample of 936 early-type galaxies  located in 48 low-z regular galaxy clusters with  $M_{200}\geq 10^{14}~ M_\odot$ at $z< 0.1$. 
We examine variations in the Kormendy relation (KR) according to their location in
the projected phase space (PPS) of the clusters.  We have used a combination of Bayesian statistical methods
to identify possible differences between the fitted relations.
Our results indicate that the overall KR is better fitted when we take into account the information about  PPS regions.
We also find that objects with time since infall $\geq 6.5$ Gyr have a
significant statistical difference of the KR coefficients relative to objects that are more recent in the cluster environment.
We show that giant central  ellipticals are responsible for tilting the KR relation towards smaller slopes. These galaxies  present a late growth  probably due to cumulative preprocessing during infall, plus cannibalism and accretion of smaller stripped objects near the center of the clusters.

\end{abstract}




\section{Introduction}
\label{section1}

The Kormendy relation (KR) is a relation between the effective radius $R_e$ of Early-Type Galaxies (ETGs) and the surface brightness $\mu_e$ at that radius \citep{kormendy1977brightness}. It is a projection of the Fundamental Plane of galaxies, a three-dimensional parameter
space correlating, in addition to $R_e$ and $\mu_e$, the velocity dispersion $\sigma$ of galaxies \citep{dressler1987spectroscopy,djorgovski1987fundamental,bender1992dynamically}.
The KR is a scaling relation indicating that more luminous objects are larger and have a lower characteristic surface
brightness.

ETGs encompass objects that normally ceased their star formation,
have red colors, small amounts of cold gas
and dust, and which correspond morphologically to ellipticals  and
lenticulars \citep[e.g.][]{kauffmann2004environmental,blanton2009physical}. Some studies indicate that the formation of massive ETGs has occurred
in two phases \citep[e.g.][]{oser2010two}. At an early stage,
the gas collapses into dark matter halos and forms stars intensely for a
short time interval \citep[e.g.][]{thomas2005epochs,peng2010mass,conroy2015preventing}. The second phase involves mass accumulation through a
series of mergers \citep[e.g.][]{naab2009minor,feldmann2011hubble,johansson2012forming,huang2016characterizing} that enrich galaxies with
stars set \textit{ex-situ}, thereby increasing their size and stellar mass continuously \citep[e.g.][]{trujillo2007strong,van2010growth}.

The structural analysis of ETGs through various scaling relations
can be used to probe the evolutionary state of galaxies
and constrain galaxy formation models and the assembly history
of ETGs \citep[e.g.][]{tortorelli2018kormendy,genel2018size,kuchner2022inventory}. In particular, the KR can be used to derive luminosity and size evolution of ETGs, but
it requires homogeneous and complete data samples for clusters at different redshifts,
wavelengths, magnitude range, taking into account 
differences between fitting methods, besides measurement and systematic errors \citep[e.g.][]{ziegler1999probing,la2003invariant,kuchner2017effects,tortorelli2018kormendy}.
All of this makes it difficult to compare results from different studies.

In this work, we address this problem by studying a homogenous sample of ETGs selected from regular galaxy clusters
\citep{ribeiro2023late}. Our aim is to study the KR in subsamples defined in the projected phase space (PPS) of the clusters. The correlation between these regions and the time since infall in clusters is used to probe
environmental and evolutionary effects on the coefficients of the KR. A brief description of our data is presented in Section \ref{section2}. Our analysis and results are presented in Section \ref{section3}. We summarize and list some conclusions in Section \ref{section4}.


\section{DATA}
\label{section2}

Our  sample is selected from the extended version of FoF group catalog originally identified by \citet{B06} and contains 5352 groups with 
$N > 5$ and  $0.03 \leq z \leq 0.11$, consisting of galaxies with  absolute magnitudes $M_r\leq -20.5$, and stellar masses in the range $10.4< \log{(M_\ast/M_\odot)} < 11.9$,\footnote{We  use  the notation “$\log{x}$” as indicating the decimal logarithm of $x$.}  with median $\sim 10^{11}~ M_\odot$. The list of members for each group was defined using the ``shifting gapper" technique \citep{fadda1996observational,L09}, extending to $\sim 4$ Mpc around the  group centers defined by \citet{Lab10}. 
The groups were then subject to the virial analysis, ana\-lo\-gous to that described in \citet{G98,P05,P07,Bi06,L09}. This procedure yields estimates of velocity dispersion ($\sigma_v$), radii ($R_{500}$, $R_{200}$) and masses ($M_{500}$, $M_{200}$) for most of the groups from the FoF sample.  Our final sample contains 107 massive clusters, with at least 20 galaxies within $R_{200}$, implying systems with $M_{200}\gtrsim 10^{14}{\rm M_\odot}$ extending to $2R_{200}$.  We reduce this sample even further, demanding that the systems do not have substructures
 according to the DS test \citep{dressler1988evidence};
and with Gaussian velocity distribution, according to the HD metric \citep[see][]{2013MNRAS.434..784R,decarvalho}.
We also added morphology from the \cite{dominguez2018improving} catalog,  and
the S\'ersic indices $n$ in the $r$-band from the catalog of \cite{simard2011catalog}. 
Our final sample has 936  ETGs as objects with morphological parameter  $T< 0$, probability of being a bulge-dominated object $P_{bulge} > 0.5$, and S\'ersic index $n> 2.5$, located in 48  regular clusters.
For more details, see \cite{ribeiro2023late}.

\section{Analysis}
\label{section3}

We study our stacked sample of cluster ETGs according to their loci in the PPS, which is built by normalizing the projected clustercentric distances and velocities by the virial radius, $R_{200}$, and velocity dispersion, $\sigma_v$, respectively.
To explore the PPS, we divide cluster galaxies into four subsamples, following the classification introduced by \cite{rhee2017phase}, see also  \cite{ribeiro2023late}.
It is important to consider possible differences between our definitions and those used by \cite{rhee2017phase}, specifically our use of $R_{200}$ and their use of $R_{vir}$, which are generally not equivalent \citep[see][]{rhee2020yzics}.
We derive a first estimate of the virial mass from equation 5 of \cite{G98}. After applying the surface pressure term correction to the mass estimate, we assume a NFW profile to obtain an estimate of M$_{200}$. This is found after the interpolation (most cases) or extrapolation of the virial mass M$_V$ from R$_A$ to R$_{200}$ (R$_A$ is the radial offset of the most distant cluster member). Then, we use the definition of M$_{200}$ and a final estimate of R$_{200}$ is derived. Further details can be found in \cite{L09}.
With this definition, $R_{vir} \approx 1.36 R_{200}$ in the ${\rm \Lambda CDM}$ cosmology \citep[see][]{gill2005evolution,reiprich2013outskirts}, 
which may introduce a bias to our results. In order to assess for any systematic effects, we performed a series of tests and found that the present analysis remains robust and is not significantly impacted by this difference in the radial normalization of the clusters. 
Consequently, we have chosen to maintain PPS normalization using $R_{200}$.

\begin{figure}
 \includegraphics[width=\columnwidth ]{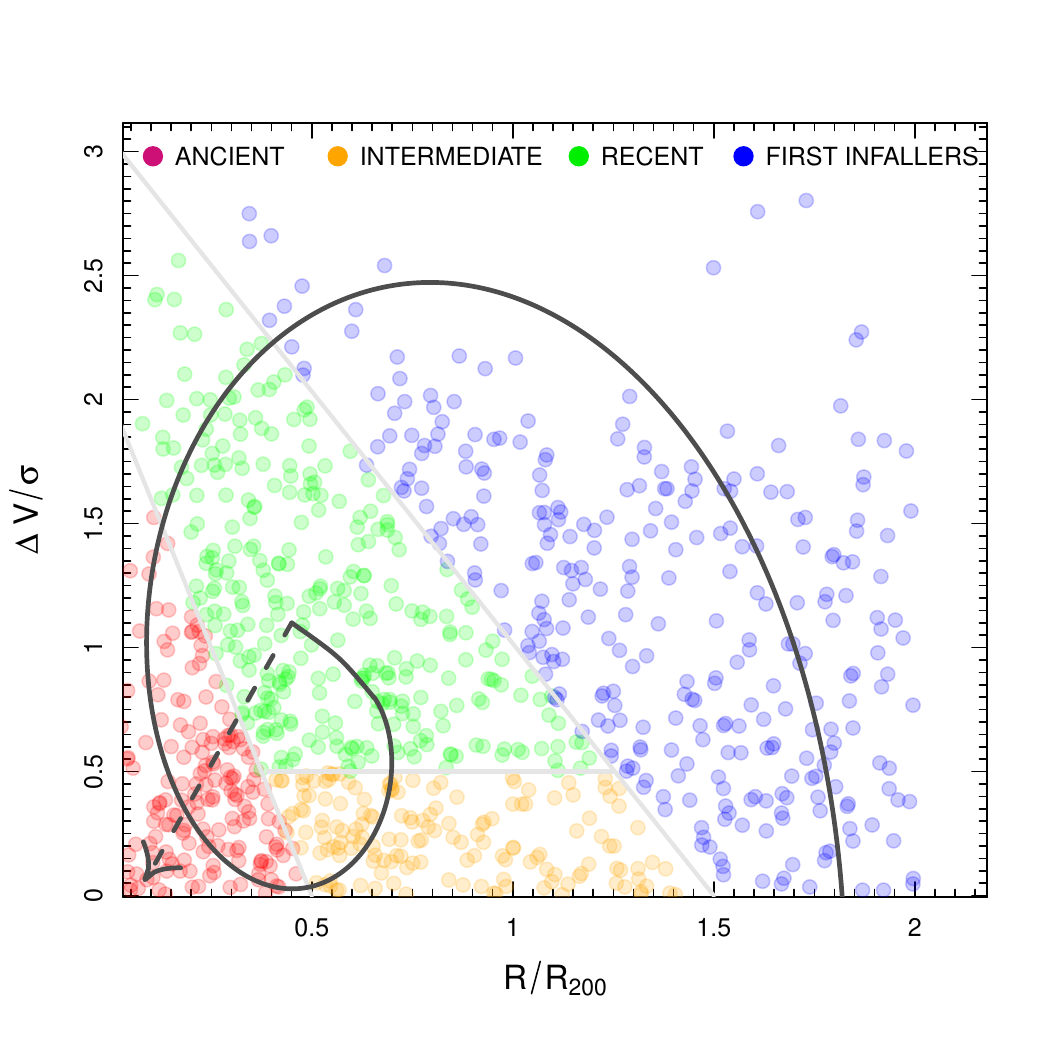}
 \vspace{-0.75cm}
 \caption{Distributions of ETGs in different PPS regions. Ancients objects are in red,
 intermediate in orange, recent in green, and the first infallers are in blue. A schematic curve illustrates the expected trajectory of 
a galaxy through the PPS.}
 \label{fig1}
\end{figure}

In Figure \ref{fig1} we see the distribution of ETG galaxies in the PPS
regions of \cite{rhee2017phase}, following the approximation of \cite{song2018redshift}.
All the ancient (171 objects), intermediate (157 objects), recent (307 objects), and
first infallers (301 objects) are shown in this figure. A schematic curve illustrates the expected trajectory of 
a galaxy through the PPS, and therefore indicates that an object in the "ancient" region requires a long time
to delve into the potential of the cluster (tipically more than 6.5 Gyr - see e.g. \citealt{rhee2017phase}).


\subsection{The Kormendy relation}

The Kormendy relation (KR) defines an observational correlation between effective radius and surface brightness of ETGs, usually
expressed as $  \mu_e = \alpha + \beta \log{R_e} $ \citep{kormendy1977brightness}. This relation provides information
on the distribution of the light profiles and the sizes of ETGs, thus it
can be used to study their
size-evolution as a function of the
environment, in particular the PPS regions.

The effective radii are calculated by 
\begin{equation}
R_e = a_{\rm deV}\sqrt{b/a}
\label{eq1}
\end{equation}

\noindent where $a_{\rm deV}$ and $b/a$ are the semimajor axis length and the axis
ratio from the de Vaucouleurs fit, respectively. We only use ETGs with $b/a > 0.3$ to avoid
edge-on objects, and $100<\sigma<420~{\rm km\;s^{-1}}$ \citep[see][]{ribeiro2023late}.
These restrictions on $\sigma$ and $b/a$ reduce the sample to 912 galaxies.
The parameters  $a_{\rm deV}$ and $b/a$  in $r$-band
are taken from the catalogs PhotObjAll and SpecObjAll of DR15
\citep{aguado2019fifteenth}.

We transformed the surface brightness $\mu_e$
at $R_e$ to $\langle \mu_e \rangle $ using the relation 
\begin{equation}
\langle \mu \rangle_e = m_r + 5\log{R_e} + 2.5\log{2\pi},
\label{eq2}
\end{equation}

\noindent recalling that we now have
\begin{equation}
\langle \mu \rangle_e = \alpha^\prime + \beta\log{R_e},
~~{\rm with}~~\alpha^\prime = \alpha - 1.4
\label{eq3}
\end{equation}

\noindent \citep[e.g.][]{graham1997some,longhetti2007kormendy}.
This is assumed for the rest of the analysis, but for simplicity we keep the nomenclature $\alpha$.

\subsection{The KR fitting - all data}

We begin our analysis by asking whether classifying the data according to PPS regions influences the result of the linear regression. We first ran the robust linear regression with M-estimator for all ETGs, regardless of the PPS region. Then, we ran the regression once more, now considering the region as a categorical predictor.

\begin{equation}
\begin{cases}
\begin{array}{rcl}
\langle\mu\rangle_e&=&\alpha + \beta \log{R_e}\\ 
\langle\mu\rangle_e&=&(\alpha + \beta \log{R_e})\cdot {\rm region} \\
\end{array}
\end{cases}
\end{equation}

\begin{figure}
\includegraphics[width=\columnwidth]{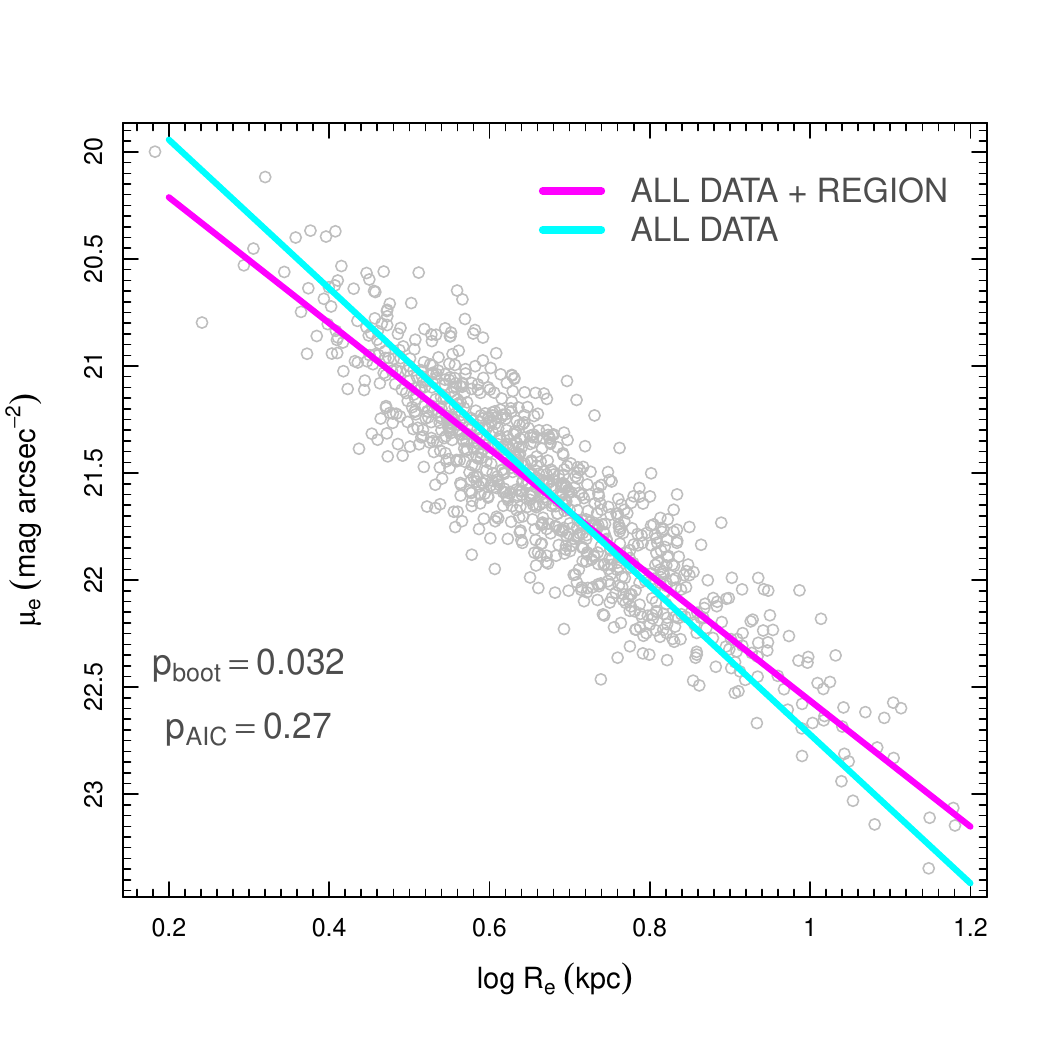}
\vspace{-0.7cm}
\caption{Linear regressions including (magenta) or not (cyan) the region as an interaction term on the slope.}
\label{fig2}
\end{figure}

\noindent Hence,  "region" is a categorical variable including the interaction effect of PPS loci on the coefficients of the KR relation.
The results are presented in Figure \ref{fig2}.
We note that both the slopes and intercepts are slightly different between the models: 
$\alpha = 19.62\pm 0.19$ and 
$\beta = 2.94\pm 0.26$ (including the PPS regions); or $\alpha = 19.24\pm 0.11$ and 
$\beta = 3.47\pm 0.15$ (not including the PPS regions). We compared the models using the Akaike Information Criterion (AIC) difference for the respective likelihoods, finding ${\rm \Delta_{AIC}=2.59}$, which provides strong support in favor of the model with influence from the PPS regions. We can assign a probability to the alternative model, without influence, as
\begin{equation}
{\rm p_{AIC}=\exp\left(\frac{-\Delta_{AIC}}{2}\right)},
\end{equation}

\noindent which provides a relative probability that the alternative model minimizes the AIC. 
We find ${\rm p_{AIC}=0.27}$, which is low but not entirely convincing. Then we resample the data 1000 times and check how many times the alternative model minimizes the AIC. This procedure results in only 3.2\% probability (${\rm p_{boot}}$) favoring the alternative model, which gives us confidence about the influence of the PPS regions to explain the KR.

\subsection{The KR fitting in PPS regions}

The KR fitting is sensitive to sample selection effects as well as to fitting procedures
\citep[e.g.][]{ziegler1999probing,la2003invariant,tortorelli2018kormendy}. 
In particular, ETGs in clusters
define the KR with a high intrinsic dispersion, probably due to measurement and systematic errors.
\citep[e.g.][]{la2003invariant, nigoche2008kormendy}
Then,
 we chose to perform a Bayesian linear regression
to recover the whole range of inferential solutions, i.e. the posterior distributions of the
KR coefficients, rather than a point estimate and a confidence interval as in classical regression. The model is defined as

\begin{equation}
   \langle\mu\rangle_e ~\sim~ \mathcal{N}\left( \alpha + \beta \log{R_e}, \sigma^2\right).
\end{equation}

\begin{figure}
\includegraphics[width=\columnwidth]{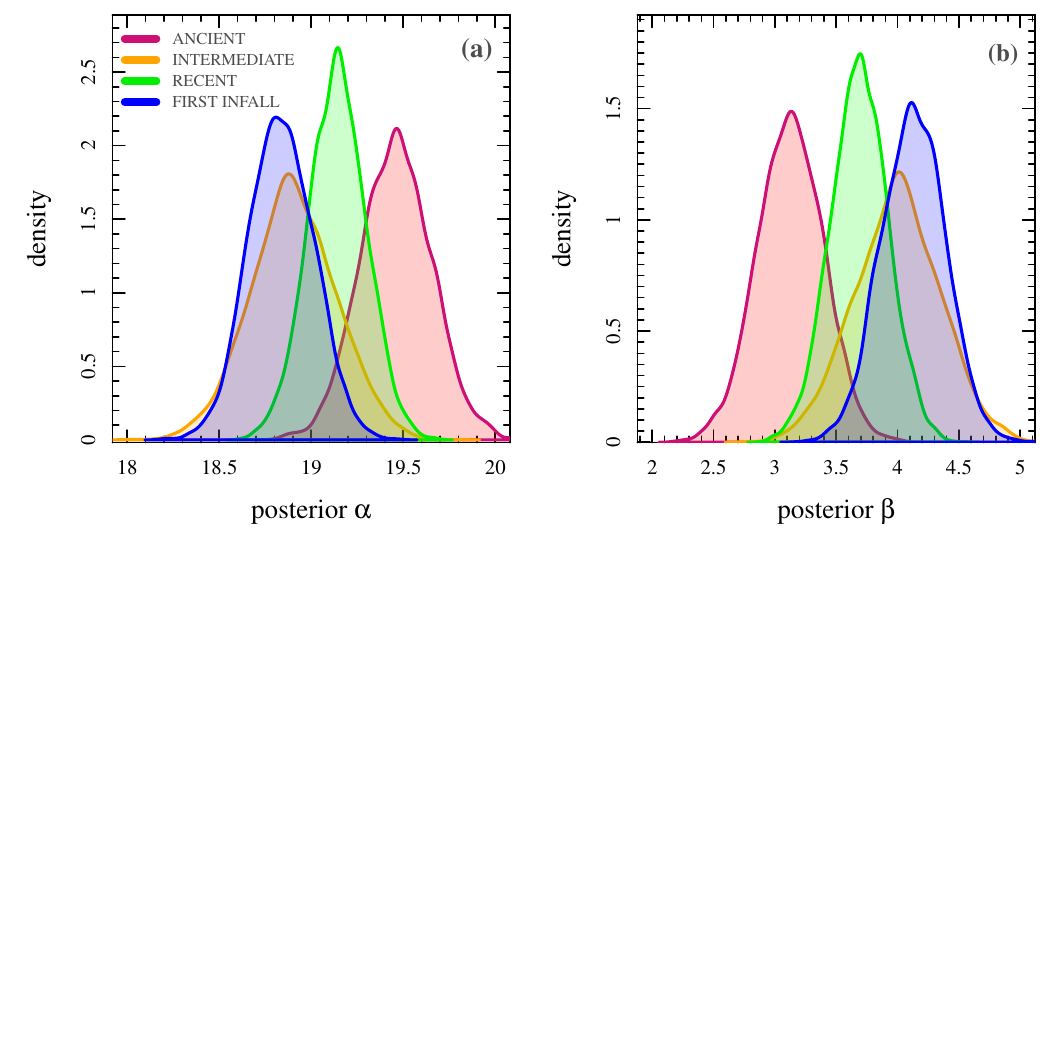}
\vspace{-4.7cm}
\caption{Posterior regression parameters from the Bayesian fits. Ancients objects are in red,
 intermediate in orange, recent in green, and the first infallers are in blue. }
\label{fig3}
\end{figure}

\begin{figure}
\includegraphics[width=\columnwidth]{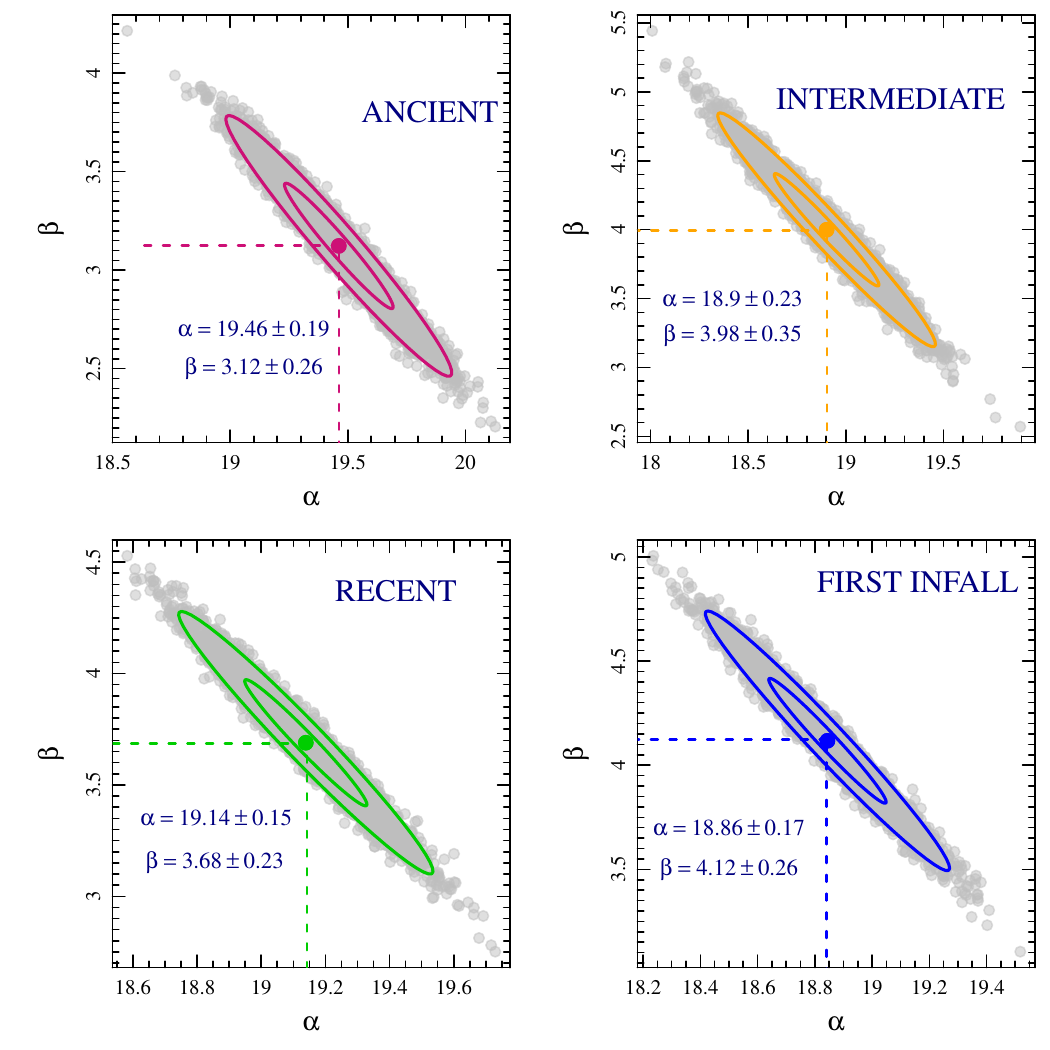}
\vspace{-0.7cm}
\caption{Confidence ellipses of parameters from the Bayesian linear fit at levels 50\% and 95\%. Ellipses for ancients objects are in red,
 intermediate in orange, recent in green, and the first infallers are in blue.}
\label{fig4}
\end{figure}

\noindent That is, we assign to the parameters a bivariate normal with noise being independent, normally distributed random variables, $\epsilon_i \sim \mathcal{N}(0,\sigma^2)$. The normal priors are taken from the best fit we found in Section 3.2. From the 
normal-normal conjugacy we know that the
posterior distributions for each regression
is also normal distributed \citep[e.g.][]{lock2020statistics}. The posteriors are presented in Figure \ref{fig3}, with the
corresponding bivariate distributions in
Figure \ref{fig4} for 5000 Markov-Chain-Monte-Carlo (MCMC) samples. The differences observed in these figures need to be tested and quantified.
For this purpose, we use Bayesian
posterior distributions for hypothesis testing. Methods include the Bayes factor two sample t-test \citep[e.g.][]{morey2011bayes}, 
the Bayesian p-value based on the density at the Maximum A Posteriori (MAP)  \citep[e.g.][]{mills2018objective}, the region of practical equivalence (ROPE) \citep[e.g.][]{kelter2022evidence} and the probability of direction (PD) \citep[e.g.][]{makowski2019bayestestr}. These methods are briefly described in Appendix A.
In all cases the quantity to be tested is the difference of slopes and intercepts ($\delta$) between two posterior distributions from the Bayesian regressions with respect to different PPS regions. We define the PPS regions as
1 - ancient; 2 - intermediate; 3 - recent; and 4 - first infall. Accordingly, the differences of the posteriors are written as $\delta_{ij}$,
where $i~{\rm and}~j$ refer to two different regions of the PPS.

\begin{figure}
\includegraphics[width=\columnwidth]{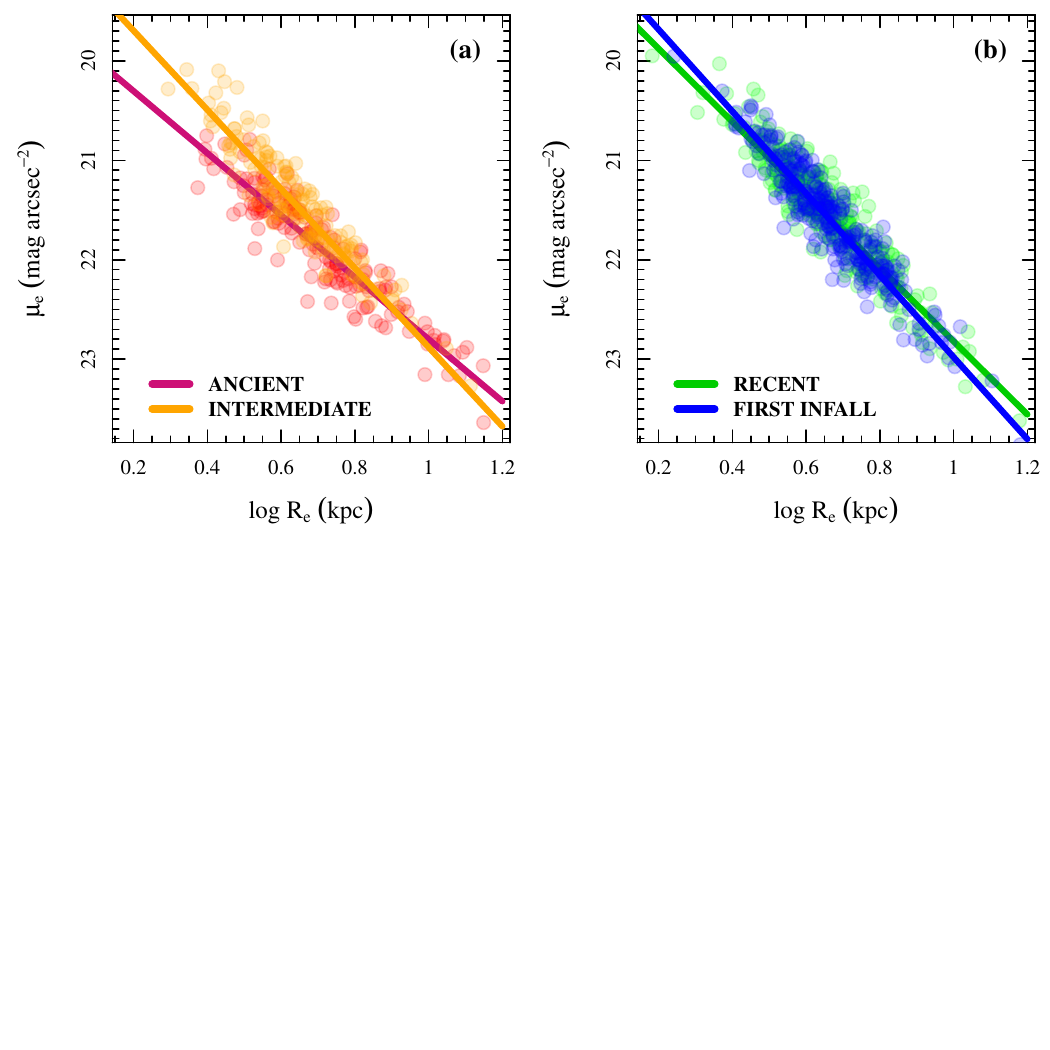}
\vspace{-4.7cm}
\caption{Linear regressions including for all subsamples. Colors indicate (a) Ancient (red) and Intermediate (orange); (b) Recent (green) and
First Infall (blue).}
\label{fig5}
\end{figure}

Our results are summarized in Tables \ref{tab1} and \ref{tab2}, where we notice an almost complete agreement between the methods. 
Red cells indicate cases in which the tests reject the null hypothesis that slopes and/or intercepts  are equivalent.
It is clear from the p-values in these tables that the most conservative test is the MAP-based one, and that all tests agree that there is no significant difference between galaxies taken from the intermediate and first infall regions.
This result is a little surprising since, considering the time since infall, these are not regions in direct sequence. In fact, the steepest Kormendy relations are precisely those of intermediate and first infall galaxies.
Objects in the recent region are in the middle, while the lowest slope comes from ancient galaxies, as we can also see
in Figure \ref{fig5}. Being conservative, we assume the results of the MAP test and interpret that the strongest and most relevant difference between the regressions is the one that distinguishes ancient objects from the other subsamples.
BCGs could be affecting this result \citep{ribeiro2023late}, but
removing them does not change the result $[(\alpha_{\rm BCG}=19.25\pm 0.19);~(\beta_{\rm BCG}=2.95\pm 0.27)]$. Likewise, removing the lenticulars we found no significant difference in the
fittings at the 95\% confidence level. This suggests an environmental-evolutionary explanation to understand the particular behavior of the KR for ancient galaxies.

\begin{table}
\centering
\caption{Results of the Bayesian tests for intercepts $\alpha$. Significant differences are highlighted in red.}
\label{tab1}
\begin{tabular}{rrrrr}
  \hline
DATA & BF & MAP & ROPE & PD \\ 
  \hline
  $\delta_{12}$  & \cellcolor{red!25}0.041 & \cellcolor{red!25}0.033 & \cellcolor{red!25}0.014 & \cellcolor{red!25}0.027   \\ 
   $\delta_{13}$ & \cellcolor{red!25}0.038 & \cellcolor{red!25}0.045&  \cellcolor{red!25}0.034 & \cellcolor{red!25}0.018 \\ 
   $\delta_{14}$ & \cellcolor{red!25}0.044 & \cellcolor{red!25}0.047  & \cellcolor{red!25}0.048 &\cellcolor{red!25}0.021   \\ 
   $\delta_{23}$ & \cellcolor{red!25}0.032 & 0.238&   0.189& \cellcolor{red!25}0.037 \\ 
   $\delta_{24}$ & 0.278&  0.333 &  0.534 & 0.199\\
   $\delta_{34}$ & \cellcolor{red!25}0.048& 0.233 & \cellcolor{red!25}0.047 & \cellcolor{red!25}0.036 \\
   \hline
\end{tabular}
\end{table}

\begin{table}
\centering
\caption{Results of the Bayesian tests for slopes $\beta$. Significant differences are highlighted in red.}
\label{tab2}
\begin{tabular}{rrrrr}
  \hline
DATA & BF & MAP & ROPE & PD \\ 
  \hline
  $\delta_{12}$& \cellcolor{red!25}0.001 & \cellcolor{red!25}0.034 & \cellcolor{red!25}0.034&  \cellcolor{red!25}0.038  \\ 
   $\delta_{13}$  & \cellcolor{red!25}0.022 & 0.289 & \cellcolor{red!25}0.023  & \cellcolor{red!25}0.047  \\ 
   $\delta_{14}$  & \cellcolor{red!25}0.019& \cellcolor{red!25}0.044& \cellcolor{red!25}0.034 &  \cellcolor{red!25}0.028\\ 
   $\delta_{23}$  & \cellcolor{red!25}0.048& 0.358 & \cellcolor{red!25}0.037  & \cellcolor{red!25}0.043  \\ 
   $\delta_{24}$  & 0.342& 0.457& 0.212 & 0.327 \\
   $\delta_{34}$  & \cellcolor{red!25}0.044 & 0.388& \cellcolor{red!25}0.039 & \cellcolor{red!25}0.047 \\
   \hline
\end{tabular}
\end{table}

\subsection{Mass-radius relation}

One can study changes in  the KR by considering the evolution of luminosity at a certain $R_e$ \citep[e.g.][]{la2003invariant}, or through the relative growth of the size of objects in relation to their stellar mass \citep[e.g.][]{van2010growth,van20143d}. In this work, we explore this second approach.  Galaxy size is a property that is found to significantly vary with galaxy mass, star-formation activity, and redshift \citep[e.g.][]{shen2003size,genel2018size}. In this context, the mass-radius relationship is of fundamental importance for probing changes in the KR. Usually, it  is assumed to be a power law \citep[e.g.][]{shen2003size,van20143d}, describing the growth
of $R_e$ with the stellar (or total) mass.
 According to \cite{van2010growth} the variation of the effective radius $R_e$ with $M_\ast$ is given by

 \begin{equation}
     {d\log{(R_e)}\over d\log{(M_\ast)}}
     \approx 3.56 \log{(n+3.09)} - 1.22.
 \end{equation}

\begin{figure}
\includegraphics[width=\columnwidth]{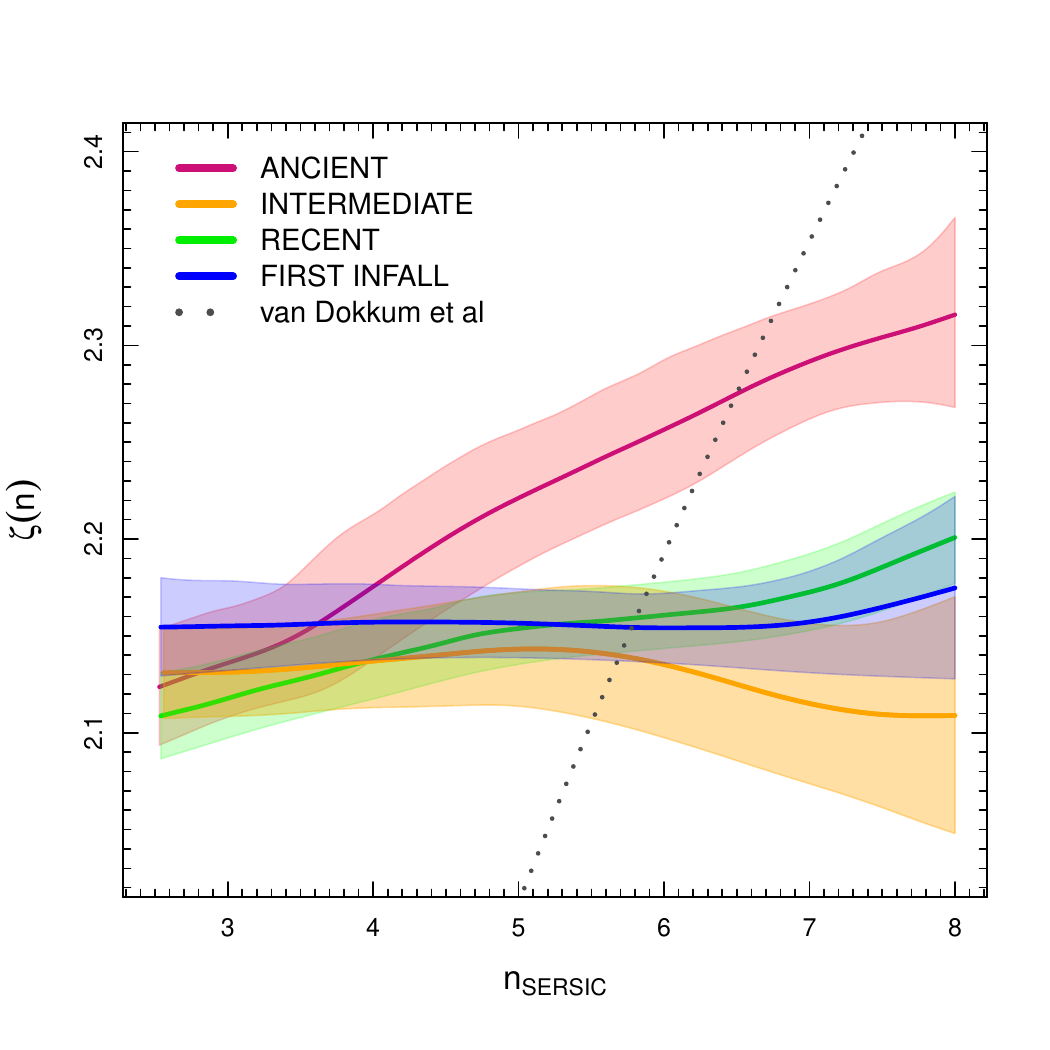}
\vspace{-1.0cm}
\caption{Growth of $R_e$ with stellar mass for all sub-samples. Colors indicate: Ancient (red); Intermediate (orange);  Recent (green); and First Infall (blue). The gray dotted line corresponds to the  expected relation of
$\zeta(n)$, while shaded areas show the confidence intervals calculated from 1000 bootstrap
realizations.}
\label{fig6}
\end{figure}

\noindent We call this variation ${\rm \zeta(n)}$, and we used it to investigate possible differences between our subsamples.
The idea to be tested is whether the time since infall promotes different behaviors for
the ETG in clusters. Using the interpolation of lagged and iterated differences of $R_e$
and $M_\ast$ we compute the smoothed behaviour
of ${\rm \zeta(n)}$, presented in Figure \ref{fig6}. In this figure, we highlight three points: (i) ancient objects present a
significant growth in size with respect the
stellar mass; (ii) all the remaining objects present a nearly constant growth, a significant deviation from the power law expectation; and (iii) the  \cite{van2010growth} growth line is steeper than the ancient behaviour.

This result clearly indicates a stronger
environmental effect on galaxies longer within clusters. According to \cite{hilz2013minor},
a key factor to understand the inside-out growth scenario is the mass-ratio of merging
galaxies, such that for minor mergers, the sizes grow significantly faster and the profile shapes change more rapidly. \cite{ribeiro2023late} found
a significant change of both size and shape (given by the S\'ersic index) for ancient satellites, suggesting the relevance of possible mergers. 

It is important to note that  
major mergers are supposed not to occur once galaxies have established themselves in the ancient region, an environment unsuitable for such mergings. The process takes place over time across the other regions, possibly through preprocessing in infall groups
\citep[see e.g.][]{lopes2024role}. On the other hand,
it is known that 
central galaxies  gradually
swallows their smaller companions as dynamical friction brings them to the cluster center, growing both
in size and mass, while losing velocity dispersion \citep[e.g.][]{goto2005velocity}. The growth
of $R_e$ with the stellar  mass for ancient ETGs seen in Figure \ref{fig6} 
 aligns with the likelihood of minor mergers and cannibalism, as supported by the aforementioned dynamical friction in the central regions of clusters.
 In fact, at least 30\% of the mass accreted by central galaxies since $z \approx 1$ is  expected to come from  mergers  and accretion from stripped satellites \citep[e.g][]{laporte2013growth, nipoti2018accretion,ragone2018bcg,chu2021physical}. 
This sets a lookback time of $\approx$7.6 Gyr, which is close to the lower limit of the time since the infall for the ancient objects \citep{rhee2017phase}.  





\section{DISCUSSION}

The Kormendy relation  links the effective surface brightness of ETGs to their effective radius. Our findings indicate a relevant role for the environment in establishing this relationship. Basically, we see that objects in the ancient region of the clusters present a tilt towards smaller slopes of the linear fit.
To better understand this result, we split the sample of ancient ETGs into BCGs, satellites with low mass ($M_\ast < 10^{11}\; M_\odot$ -- LM ANC SAT), and satellites with high mass  ($M_\ast \geq 10^{11}\; M_\odot$ -- HM ANC SAT). 
\cite{ribeiro2023late} verified a change in the behavior of the structural properties of galaxies close to this value,  which is also approximately the median stellar mass of the entire sample. The subsample of BCGs is composed
of the 48 most luminous ancient cluster ETGs (each taken from a
cluster) having $M_r < -21.4$ and $M_\ast \geq 10^{11}\; M_\odot$ \citep[see][]{ribeiro2023late}. The LM and HM ancient satellites samples contain 56 and 67 objects, respectively. We also consider a sample of isolated early-type
galaxies taken from the list of 1-member groups of the catalog of \cite{yang2007galaxy}. This field sample, composed of 6670 objects, is
defined with the same criteria (except membership) used to select the
cluster ETGs \citep[see also][]{ribeiro2023late}.

\begin{figure}
\includegraphics[width=\columnwidth]{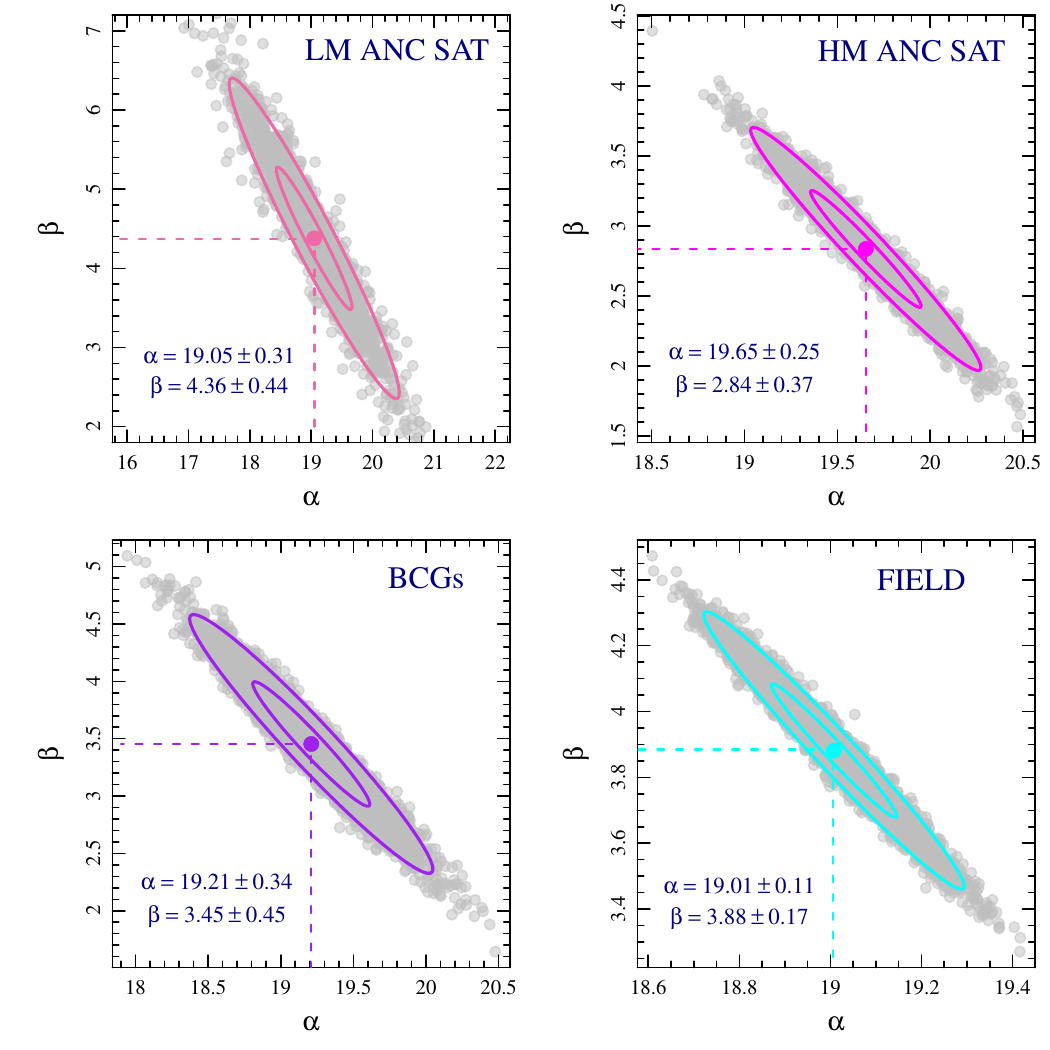}
\vspace{-0.8cm}
\caption{Confidence ellipses of parameters from the Bayesian linear fit at levels 50\% and 95\%. Colors indicate
Low Mass Ancient Satellites (pink), High Mass Ancient Satellites (magenta), BCGs (purple), and Field galaxies (cyan).}
\label{fig7}
\end{figure}

We compared the linear coefficients of the KR relation for each subsample, following the procedure described in Section 3.3. The confidence
ellipses of the parameters from the Bayesian linear linear fit  are presented in
Figure \ref{fig7}, where the bivariate distributions correspond to 5000 Markov-Chain-Monte-Carlo (MCMC) samples. Using the methods described in Section 3.3 and 
Appendix A, we find no significant difference between
the intercepts of the subsamples. The result of the slope comparison is presented in Table \ref{tab3}.
Numbers 1 to 4 indicate the datasets LM SAT, HM SAT, BCGs, and FIELD galaxies, respectively.

The strongest result is the statistical evidence of a smaller slope for HM ancient satellites with respect to the other subsamples, see Figure \ref{fig7}. For HM ANC SAT, $\sim$90\% of
objects have $T< -2$, so they are dominated by massive ellipticals. Our result indicates that this population  is likely responsible for tilting the KR relation towards smaller slopes. On the other hand, the comparison for BCGs does not suggest a significant difference in relation to the other subsamples, except the HM ANC SAT. 
This result does not agree with \cite{samir2020fundamental} who find smaller slopes for BCGs in relation to the field. This disagreement may be related to differences in the definition of BCGs, as well as to uncertainties in the brightness profiles of these objects, which are difficult to map due to the presence of many other galaxies close to them. In fact, for mergers occurring at late times, BCGs mainly accrete mass into their extended outskirts, beyond the observational photometry apertures
\citep[e.g.][]{inagaki2015stellar}.  The multiple cores in BCGs at $z\lesssim 0.1$  also suggests a complex stellar mass growth of these objects
over the past 4.5 Gyr \citep{hsu2022sdss} . This growth could be affecting the slope of the KR relation, but our results do not indicate this, either due to problems in determining their structural properties, or due to some mechanism regulating the growth in brightness and size of these galaxies.
Differently, the slope variation for the  HM ANC SAT is significant, 
which may be associated with both  late accretion in the central region and the history of mergers in infall groups over the cluster evolution \citep[e.g.][]{lopes2024role}. It is also important to mention that field galaxies do not present significantly different coefficients from the intermediate, recent and first infall samples, reinforcing the idea of evolution in ancient galaxies.

\begin{table}
\centering
\caption{Results of the Bayesian tests for slopes $\beta$ of subsamples of LM SAT, HM SAT, BCGs, and FIELD galaxies.}
\label{tab3}
\begin{tabular}{rrrrr}
  \hline
DATA & BF & MAP & ROPE & PD \\ 
  \hline
  $\delta_{12}$& \cellcolor{red!25}0.001 & \cellcolor{red!25}0.003 & \cellcolor{red!25}0.004&  \cellcolor{red!25}0.005  \\ 
   $\delta_{13}$  & \cellcolor{red!25}0.047 & 0.115 & \cellcolor{red!25}0.043  & 0.236  \\ 
   $\delta_{14}$  & 0.298 & 0.322& 0.145 &  0.379\\ 
   $\delta_{23}$  & \cellcolor{red!25}0.009& 0.188 & \cellcolor{red!25}0.027  & \cellcolor{red!25}0.039 \\ 
   $\delta_{24}$  & \cellcolor{red!25}0.008& \cellcolor{red!25}0.016& \cellcolor{red!25}0.031 & \cellcolor{red!25}0.018 \\
   $\delta_{34}$  & 0.126 & 0.333& 0.284 & 0.227 \\
   \hline
\end{tabular}
\end{table}

Therefore, our work is consistent with a scenario where giant ellipticals show a late growth, $z \lesssim 0.1$, either through infall group interactions or through interactions and accretion promoted by dynamic friction in the central region of the clusters. The combination of these evolutionary channels produce a significant effect on the Kormendy relation, indicating that it can be used as an important indicator of the late evolution of cluster galaxies.

\section{Conclusions}
\label{section4}

The study of  ETGs is important to improve our understanding of the structure
formation and the galaxy–environment connection. 
In this work, we investigated the KR of cluster ETGs according to their 
loci in the PPS. We have used a combination of statistical methods
to identify possible differences between the fitted KR. From our analysis we can 
conclude that:

\begin{itemize}
\item The KR is  is better fitted when we take into account information about the PPS regions.
\vspace{0.2cm}

\item There is a significant statistical difference between the KR coefficients of the ancient ETGs in relation to the others.
\vspace{0.2cm}

\item The longer time within host clusters and the cumulative effect of mergers (in infall groups and/or in the central regions due to dynamical friction) plus 
accretion of stripped galaxies are the likely causes for the difference between the KR through the PPS regions. 
The relative importance of these  mechanisms deserves a further study, to be presented in a forthcoming paper.
\vspace{0.2cm}

\item HM ancient satellites are the component that most contributes to tilting the KR relation.
\vspace{0.2cm}

\item A better determination of the brightness profile of BCGs and a larger sample of these objects can help to better understand the slope of the KR relation in the ancient region.

\end{itemize}

 A possible caveat to this work is that we have not  considered the effects of the velocity dispersion on the intrinsic scattering of the KR, a point to be examined in the future, in a complete study of the fundamental plane.

\section*{Acknowledgements}
The authors thank the referee for 
useful suggestions.
ALBR thanks the support of CNPq, grant 316317/2021-7 and FAPESB INFRA PIE 0013/2016. RSN thanks the financial support from CNPq, grant 301132/2020-8.
PAAL thanks the support of CNPq, grants 433938/2018-8 e 312460/2021-0. CCD thanks the support by the Coordena\c c\~ao de Aperfei\c coamento de Pessoal de N\'{\i}vel Superior - Brasil (CAPES) - Finance Code 001, the Programa Institucional de Internacionaliza\c c\~ao (PrInt - CAPES), and the Brazilian Space Agency (AEB) for the funding (PO 20VB.0009). MHSF  and FRMN thank the financial support by the Coordenação de Aperfeiçoamento de Pessoal de Nível Superior - Brasil (CAPES) - Finance Code 001. BGA thanks the support of PROBOL-UESC.

This research has made use of the SAO/NASA Astrophysics
Data System, the NASA/IPAC Extragalactic Database (NED) and
the ESA Sky tool (sky.esa.int/). Funding for the SDSS and SDSS-II
was provided by the Alfred P. Sloan Foundation, the Participating Institutions, the National Science Foundation, the U.S. Department of Energy, the National Aeronautics and Space Administration, the Japanese Monbukagakusho, the Max Planck Society, and the Higher Education Funding Council for England.

\section*{Data Availability}

The data that support the findings of this study are available on request from the corresponding author, A.L.B.R.,  upon reasonable request.



\bibliographystyle{mnras}
\bibliography{refs}



\appendix

\section{Bayesian hipothesis testing}
\begin{figure}
\includegraphics[width=\columnwidth]{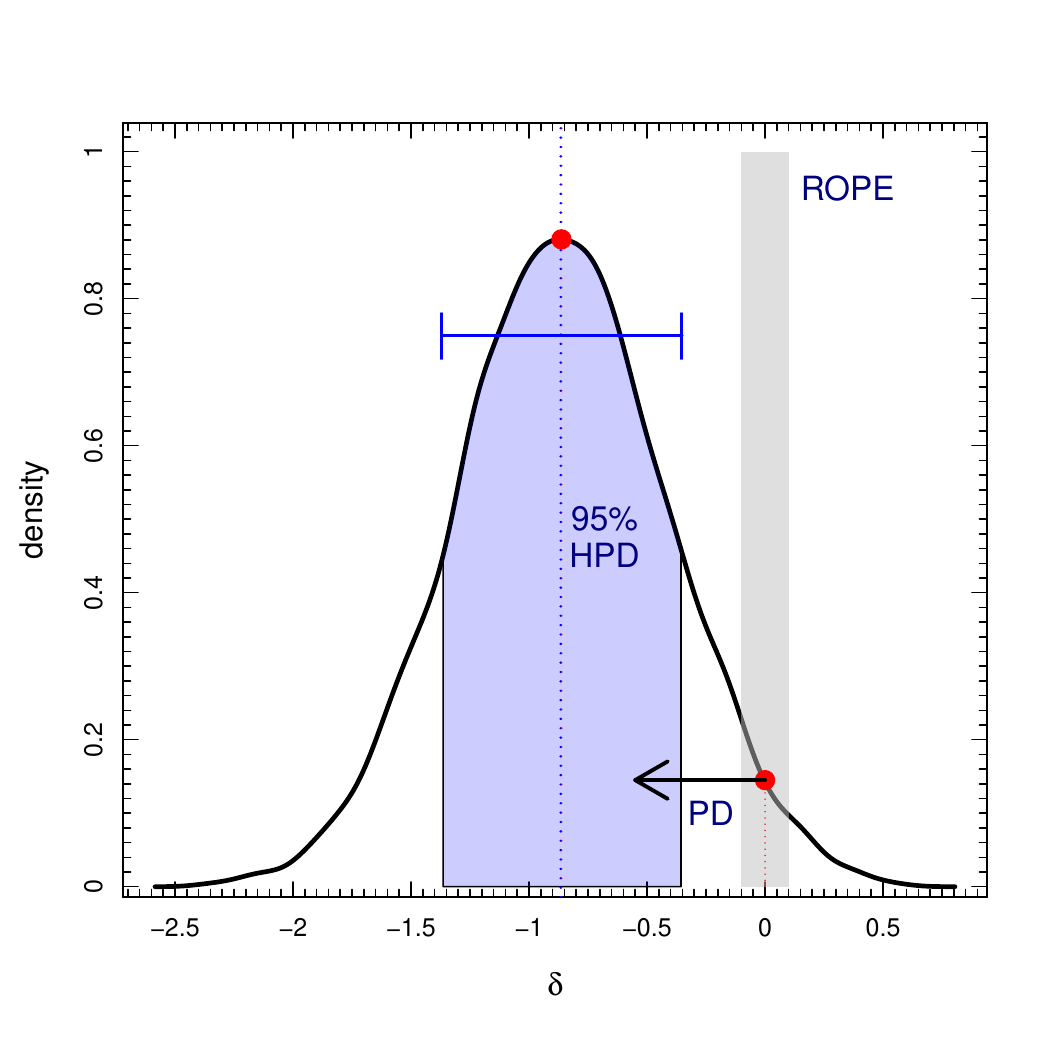}
\vspace{-1.0cm}
\caption{Figure illustrating the most important aspects of the Bayesian methods used in this work. The gray
shade area shows the ROPE, while the blue shaded area corresponds to the posterior probability
 inside the 95\% HPD. For the likelihood ratio, points in red are showed, and for the
 computation of the PD area, the arrow indicates  the lower limit of the density distribution integral, which extends to $-\infty$.}
\label{fig_ape}
\end{figure}

 A statistical hypothesis is a hypothesis about a particular model parameter or a set of model parameters. Bayesian approaches to hypotheses testing are  foremost concerned, not with categorical decisions, but with quantifying evidence in favor or against the hypothesis in question. In this Appendix we briefly describe
 the methods used in Section 3.3. Here, in 
 all cases the quantity to be tested is the difference of slopes and intercepts ($\delta$) between two posterior distributions from the Bayesian regressions, with null hypothesis $\delta=0$. The tests were done using functions contained in
 two R libraries: {\bf bayesfactor} and
 {\bf bayestestR}. In Figure \ref{fig_ape}illustrates the most important elements of the Bayesian methods used in this work.

\subsection{Bayes Factor (BF) t-test}

The Bayes factor is a wide used index 
for testing a hypothesis in the Bayesian approach
\citep[see e.g.][]{jeffreys1998theory}. To test the null hypothesis $\delta=0$,
 a Cauchy prior with location zero and scale 
 $r = 1/\sqrt{2}$ is used. For the variance
 $\sigma^2$,  Jeffreys’s prior is used:
 $p(\sigma^2) \propto 1/\sigma^2$. In this example, we are interested in comparing the null model $H_0$, which posits that the difference between the posteriors is zero, to the alternative hypothesis $H_1$, which assigns $\delta$ the Cauchy prior. To compare the null model and the alternative model, we can compute the Bayes factor 
 ${\rm BF_{10}}$, that is, the Bayes factor which quantifies how much more likely the data are under the alternative versus the null hypothesis. 
 ${\rm BF_{10}}$ is the Bayes factor giving evidence for $H_1$ over $H_0$. The two-sample Bayesian t-test is  done with the relationship between the hypotheses used in ${\rm BF_{10}}$  and the Student's $t$ given by

 \begin{equation}
     H_0: ~ t~\sim ~ T_\nu ,~~~~{\rm and}
 \end{equation}
 \vspace{-0.8cm}
\begin{equation}
    H_1: {t\over \sqrt{1+\sigma_\delta^2 N}} ~\sim ~
    T_\nu \left({\delta\over \sqrt{1/N + \sigma_\delta^2}}\right)
\end{equation}

\noindent where $T_\nu$ represents the standard (central) $T$ distribution with $\nu$ degrees of freedom, and  $T_\nu(\theta)$ is the non-central 
$T$ distribution with argument $\theta$
\citep[e.g.][]{gonen2005bayesian}.

\subsection{Maximum A Posteriori (MAP)}

The MAX-based p-avalue is
defined as the ratio of the posterior densities null value $p(\theta_0)$ and the posterior densities maximum a posteriori (MAP) value:

\begin{equation}
    p_{\rm MAP} = {p(\theta_0|x)\over \arg{\underset{\theta\in \Theta}\max{P(\theta|x})}}
\end{equation}

\noindent where $\theta_0=\delta=0$. The test is
based on the likelihood ratio used in the Neyman-Pearson theory \citep{neyman1933ix}. The denominator is not maximising the likelihood over the alternative hypothesis $H_1$ like in a traditional Neyman-Pearson test, but instead the posterior density $p(\theta|x)$ is maximised.
The MAP-based p-value is the ratio between the
posterior distribution's value at the null value
$\theta=0$ and the maximum of the distribution
\citep[][]{mills2018objective}.

\subsection{Region of Practical Equivalence (ROPE)}

The general idea of this test is to establish a region of practical
equivalence around the null value $\theta_0$ of the
null hypothesis $H_0:\theta=\theta_0$, which express the range of parameter values that are equivalent to the null value for practical purposes \citep{kruschke2018bayesian}. To test hypotheses via the ROPE we need do find the highest posterior density (HPD) interval, which is basically the shortest interval on a posterior density for some given confidence level.
\cite{kruschke2018rejecting} proposed the following
decision rule:

\begin{itemize}
    \item Reject the null value $\theta_0$ specified by $H_0: \theta=\theta_0~ (\delta=0)$, if the 95\% HPD falls entirely outside de ROPE.
    \vspace{0.25cm}
    
    \item Accept the null value, if the 95\% HPD falls enterily inside the ROPE.
 \end{itemize}

Hence, if the 95\% HPD falls entirely inside the ROPE, the
parameter value is located inside the ROPE with at least
95\% posterior probability. As a consequence, it is practically
equivalent to the null value $\theta_0$ and it is reasonable to accept $H_0$ \citep[see][]{makowski2019bayestestr}.

\subsection{Probability of Direction (PD)}

The probability of direction (PD) is  dvan2010growthefined as
the proportion of the posterior distribution that is of the median’s sign,is a measure of effect existence representing the certainty with which an effect is positive or negative. Therefore, the PD is simply the proportion of the posterior
probability density which is of the median's sign.
Based on the PD, one can test the null hypothesis
$H_0:\delta =0$ by requiring a specified amount of posterior probability density to be strictly positive or negative. Usually, more than 95\% of
the posterior can be used as threshold for deciding
between $H_0$ and $H_1$. The two-side p-value is obtained as

\begin{equation}
    p_{\rm PD} = 2\cdot (1 - {\rm PD}).
\end{equation}

\noindent \citep[see][]{makowski2019bayestestr}.

\bsp	
\label{lastpage}
\end{document}